\documentclass[twocolumn,superscriptaddress,preprintnumbers,pre]{revtex4}
\usepackage{amsmath}
\usepackage{graphicx}

\newcommand\be{\begin{equation}}
\newcommand\ee{\end{equation}}                                                                                
\newcommand\ba{\begin{eqnarray}}
\newcommand\ea{\end{eqnarray}}

\begin{document}

\title{
Nonlinear propagation of light in structured media:\\
generalized unidirectional pulse propagation equations
}

\author{J. Andreasen}
\affiliation{
  College of Optical Sciences, University of Arizona,
  Tucson, AZ 85721, U.S.A.
}
\author{M. Kolesik}
\affiliation{
  College of Optical Sciences, University of Arizona,
  Tucson, AZ 85721, U.S.A.
}
\affiliation{
  Department of Physics, Constantine the Philosopher University, 
  Nitra, Slovakia
}

\date{\today}

\begin{abstract}
Unidirectional pulse propagation equations [UPPE, Phys. Rev. E \textbf{70}, 036604 (2004)]  
have provided a theoretical underpinning for computer-aided investigations into 
dynamics of high-power ultrashort laser pulses and have been successfully 
utilized for almost a decade. 
Unfortunately, they are restricted to applications in bulk media or, 
with additional approximations, to simple waveguide geometries in which only a few 
guided modes can approximate the propagating waveform. 
The purpose of this work is to generalize the directional pulse propagation equations 
to structures characterized by strong refractive index differences and material interfaces.
We also outline a numerical solution framework that draws on the combination of the
bulk-media UPPE method with single-frequency beam-propagation techniques. 
\end{abstract}

\maketitle

\section{Introduction}

Computer simulations in the field of nonlinear optics have been playing an important role in understanding
ever more extreme regimes in light-matter interactions.
Dynamics of ultrashort, high-power laser pulses is one particular field, which motivated significant effort and
concomitant progress in numerical methods designed for optics at femtosecond time scales. 
One can say that optical filamentation played the role of a catalyst for the development of a number of 
pulse propagation models, which made detailed studies of extremely nonlinear regimes a possibility. 
However, the most accurate pulse propagation models remain restricted to bulk media, both gaseous and 
condensed, while waveguiding structures have to be treated with more significant approximations.

The purpose of this article is to put forward a theoretical framework, which will allow the implementation of
simulators capable of handling pulse propagation regimes characterized by the following four attributes:
\begin{itemize}
\item[A)] Structures with strong refractive index contrasts.

\item[B)] Directional long-distance wave propagation.

\item[C)] Strong waveform reshaping, both in time and space.

\item[D)] Extreme spectral dynamics, with resulting spectra often
          encompassing more than an octave in frequency.
\end{itemize}
This combination is rather difficult to handle numerically. 
For example, there exists a wealth of work 
(e.g., \cite{roeyJOSA81,chungJQE90,raoPTL99,scarmozzinoJSTQE00,saitohJQE02})
utilizing the beam propagation method (BPM), which are designed for regimes A and B, and can incorporate 
certain weak nonlinearities \cite{hendowAO86}, but are restricted to narrow spectral regimes. 
Time-domain beam propagation methods have been developed (e.g. \cite{Shibayama:05,Masoudi:11}), 
though they concentrate mainly on linear regimes.
Direct Maxwell's equations solvers \cite{Taflove2005,greeneOE06,schmitzJOSAB12} are well-suited to 
regimes A, C, and D, but are prohibitively expensive for simulating long-distance pulse propagation. 
Simulators based on the Unidirectional Pulse Propagation Equation (UPPE) \cite{Kolesik2002,kolesikPRE04} 
and other types of one-way propagation equations 
\cite{Brabec97,Crosignani1997,Trippenbach1997,Trippenbach1998,Geissler99,porras_propagation_1999,Husakou01,Kinsler2003,kinsler_theory_2005,kinslerPRA10}
can cope with attributes B, C, and D, but require additional approximations to simulate waveguiding structures, 
such as hollow-slab leaky waveguides \cite{Chen:11} or nonlinear nano-waveguides 
\cite{Foster:08,Lamont:08,Halir:12}. 
In other words, methods suitable for the combination A+B+C+D have yet to be developed. 

In this paper, we present a step in this direction and describe a generalization of the UPPE, which can be applied 
to nonlinear structured media with strong differences between refractive indices of the constituent materials. 
Departing from the wave equations, we derive an auxiliary evolution system.
This is used to find projection operators that extract forward and backward propagating components of the field
from an arbitrary optical field waveform. 
These operators transform the auxiliary system into a coupled forward-backward pulse evolution system
that is exact and accounts for structured media. 

Key to the approach we present is that the pulse evolution equations
are cast in a form which makes it possible to combine proven numerical methods.
More specifically, nonlinear interactions can be treated by ordinary differential equation (ODE) libraries, 
the same way it has been done with UPPE-based simulators \cite{Guide11}; 
the linear propagator can be treated by tapping the rich knowledge base of BPM, and in particular the techniques 
developed for wide-angle BPM (WA-BPM) (see Refs. \cite{hadleyOL92,Hadley:92} for early formulation, and Refs.
\cite{Lu02beampropagation,Le:09} for examples of various Pad\'e approximated propagators).

The remainder of this paper is organized as follows. 
First, in Sec. \ref{sec:uppe}, we give a brief summary of the UPPE model.
In Sec. \ref{sec:dppe}, this model is generalized and a coupled forward-backward pulse evolution system derived.
A unidirectional propagation approximation is applied and the resulting equation transformed into a form 
analogous to a bulk-medium UPPE. 
Considering a homogeneous medium finds the generalized equation reduces back to UPPE.
Section \ref{sec:numeric} outlines a strategy for a numerical solution of the generalized system.
We summarize and discuss future directions in Sec. \ref{sec:sum}.

\section{Unidirectional Pulse Propagation Equations: Summary\label{sec:uppe}}

The main purpose of this paper is to generalize the UPPE framework.
For the reader's convenience and ease of reference, we briefly summarize the corresponding equations.

In a homogeneous medium characterized by a dielectric permittivity $\epsilon(\omega)$, 
a pair of coupled UPPEs are exact \cite{kolesikPRE04,kolesikJST12} and can be written in the form
\begin{eqnarray}
  &&\ \ \ \partial_z \vec{E}^{\perp}_{+}(k_\perp,\omega,z)=+ik_z\vec{E}^{\perp}_{+}(k_\perp,\omega,z) + \label{eq:z1}\\
  &&\sum_{s=1,2} \vec e^{\perp}_s \vec e^{\phantom{\perp}}_s \cdot \left[ \frac{i\omega^2}{2 \epsilon_0 c^2 k_z} \vec{P}(k_\perp,\omega,z)  - \frac{\omega}{2 \epsilon_0 c^2 k_z}\vec J(k_\perp,\omega,z)\right] \nonumber
\end{eqnarray}
\begin{eqnarray}
  &&\ \ \ \partial_z \vec{E}^{\perp}_{-}(k_\perp,\omega,z)=-ik_z\vec{E}^{\perp}_{-}(k_\perp,\omega,z) - \label{eq:z2}\\
  &&\sum_{s=1,2} \vec e^{\perp}_s \vec e^{\phantom{\perp}}_s \cdot \left[ \frac{i\omega^2}{2 \epsilon_0 c^2 k_z} \vec{P}(k_\perp,\omega,z) - \frac{\omega}{2 \epsilon_0 c^2 k_z}\vec J(k_\perp,\omega,z)\right] \nonumber
\end{eqnarray}
These equations describe the evolution of $\vec{E}^{\perp}_{\pm}(k_\perp,\omega,z)$,
which are the spectral (Fourier) representation of the electric field.
$k_\perp = \{k_x, k_y, 0\}$ are the transverse wave numbers
and $k_z$ is the $z-$component of the wave-vector 
\begin{equation}
  \vec{k} = \left\{k_x, k_y, k_z \equiv \sqrt{\omega^2\epsilon(\omega)/c^2 - k_x^2 - k_y^2} \right\} \ ,
\end{equation}
which satisfies the dispersion relation $k^2 = \omega^2 n^2(\omega)/c^2$.
The two polarization vectors $\vec e_s(k_\perp,\omega)$ are orthogonal to $\vec k$ and to each other, 
but otherwise can be chosen freely. 
The superscript $\perp$ denotes the transverse part (i.e., $x,y$) of the corresponding vector. 
Equations (\ref{eq:z1}) and (\ref{eq:z2}) are mutually coupled through the nonlinear medium polarization 
$\vec{P}(k_\perp,\omega,z)$ and current density $\vec J(k_\perp,\omega,z)$. 
These responses are functionals of the electric field. 
They are normally specified in the real space and time representation:
\[
\vec{P}(r_\perp,t) = \vec{P}(\{\vec E(x,y,t)\}) \  ,  \ 
\vec{J}(r_\perp,t) = \vec{J}(\{\vec E(x,y,t)\}) \ .
\]

It has to be emphasized that the system of Eqs. (\ref{eq:z1}) and (\ref{eq:z2}) is exact and together with
the $\nabla \cdot \vec D$ equation (which can be used to obtain the $z$ component of the field if needed),
is equivalent to Maxwell's equations. 
However, as with direct Maxwell's equations solvers, it would be difficult to solve in its entirety, 
i.e., including forward and backward propagating waves.
In practice, the unidirectional propagation approximation is assumed, and the medium response is 
calculated solely from the forward propagating waveform:
\begin{equation}
  \vec P(\vec E), \vec J(\vec E) \to \vec P(\vec E_+), \vec J(\vec E_+)\ . \label{eq:approxz} 
\end{equation}
Under this approximation the system reduces to a single UPPE, Eq. (\ref{eq:z1}). 
For details of the numerical solution, the reader is referred to Ref. \cite{Guide11}.
Here we only point out that the native representation suitable for numerical implementation 
relies on spectral amplitudes $\vec{A}_{s,+}(k_\perp,\omega,z)$, which only change with $z$ 
due to nonlinear interactions with the medium. 
They are related to the electric field through the linear propagator
$e^{ik_z(k_x,k_y,\omega ) z}$:
\begin{equation}
  \vec{E}^{\perp}_+(k_\perp,\omega,z) = 
   \sum_{s=1,2} \vec e^{\perp}_s \vec{A}_{s,+}(k_\perp,\omega,z) e^{ik_z(k_x,k_y,\omega ) z}.
\end{equation}
The corresponding UPPE equation,
\begin{eqnarray}
  &&\partial_z {A}_{s,+}(k_\perp,\omega,z) = \frac{ \omega e^{-i k_z z}  }{2 \epsilon_0 c^2 k_z}\label{eq:As}\\
  &&\ \ \ \ \  \vec e_s \cdot [ i\omega \vec{P}(k_\perp,\omega,E_+(z)) - \vec  J(k_\perp,\omega,E_+(z))]\nonumber.
\end{eqnarray}
constitutes a large system of ODEs. 
This is the representation in which it is solved numerically. 
Because the medium response is calculated in the real time representation at each spatial point, 
spectral transforms in both directions have to be invoked multiple times when the right hand side 
of the ODE system is evaluated.

The main limitation of the UPPE approach is that it is restricted to homogeneous media.
Weakly guiding structures can be included as part of the polarization response, but geometries with 
strong material contrasts and interfaces cannot be efficiently simulated. 
Reference \cite{kolesikPRE04} shows derivation of the UPPE system for waveguiding structures, but its 
implementation requires knowledge of the full system of electromagnetic modal fields, which is impractical to
obtain even for geometries that admit exact solutions. 
Therefore, waveguiding scenarios can only be simulated under additional assumptions, which require that the
fields can be described as a superposition of a few guided or leaky modes of the structure,
whatever field configurations evolve.

The practical limitation of the UPPE, and in fact of all other uni-directional pulse propagation methods originates in
the identification of the forward and backward propagating waves resting on the usage of
a reference homogeneous medium. To elucidate this, let us consider the following example. Let $\epsilon(\omega)$
represent a chosen homogeneous background, and let $\chi(\vec r_\perp, \omega)$ be such that
$\epsilon + \chi(r_\perp)$ gives the actual permittivity of the structure. The variable part $\chi(r_\perp)$ can
be treated within the polarization term of Eq. (\ref{eq:As}), $P(r) = \epsilon_0 \chi(r_\perp) E_+(r_\perp)$.
Now note that even if $\chi$ is constant throughout space, the propagation constant of a plane
wave predicted by Eq. (\ref{eq:As}), i.e., $k_z(\omega,k_\perp) + \chi \omega^2/(2 c^2 k_z(\omega,k_\perp))$,  
is only a second order Taylor approximation to the exact plane-wave propagation constant
$\sqrt{\omega^2 [\epsilon(\omega) + \chi(\omega)]/c^2 - k_\perp^2}$. 
On the other hand, it is straightforward to show that if we retained both coupled UPPEs 
(\ref{eq:z1}) and (\ref{eq:z2}), the resulting propagation constant would be exact independently of the 
choice of the reference $\epsilon(\omega)$.
We thus see that if $\chi(r_\perp)$ varies in space, no matter
how we select the reference medium $\epsilon(\omega)$, even the linear problem is not solved exactly
by  a single UPPE.
This means that if we wish to use the uni-directional approximation for nonlinear interactions, 
we must find a way to marry it with an exact propagation description in the linear limit. This in turn implies 
that no part of the refractive index variation in space should be included in the polarization term of the propagation equation.

\section{Directional pulse propagation equations\label{sec:dppe}}

In this section, we generalize unidirectional pulse propagation equations to situations with material 
interfaces parallel to the 
propagation direction $z$ and with strong refractive index differences between materials that comprise 
the structure. 
Central to this task will be the ability to extract the true forward and backward propagating components 
of the total electromagnetic field. 
The main deviation from the method described above is that the pulse propagator native representation will be 
mixed; we will retain the spectral representation of the frequency/time dimension, 
but will use the real space representation for the transverse spatial dimensions $x,y$.

\subsection{Model of a nonlinear, structured medium}

Consider a non-magnetic, isotropic dispersive medium, with the dielectric permittivity $\epsilon(x,y,\omega)$,
which only depends on coordinates $x$ and $y$ and angular frequency $\omega$. 
We assume there are no free charges or currents. 
The constitutive relation for all media will be written in a form using polarization 
to account for all properties except the linear $\epsilon(\omega)$
\be
{\mathbf P} = {\mathbf P}(x,y,\{{\mathbf E}(x,y,t)\}) \ .
\ee
We assume that an algorithm is given that computes polarization from a given history of the electric 
field vector ${\mathbf E}(x,y,t)$ at a specified point $[x,y]$. 
The first two arguments of ${\mathbf P}$ are meant to indicate that this algorithm can depend, 
through the medium properties, on the transverse location $[x,y]$ but not on the longitudinal coordinate $z$. 
The concrete functional form of ${\mathbf P}$ is unimportant for the present purposes, but for a specific 
example, the reader can think of the instantaneous optical Kerr effect in which the local index of refraction 
changes proportionally to the square of the electric field vector. 
As the medium is isotropic, the polarization direction follows that of ${\mathbf E}$:
\ba
&&{\mathbf P}_{Kerr}(x,y,\{{\mathbf E}(x,y,t)\}) \nonumber\\
&&\ \ \ \ \  = 2\epsilon_0 \bar{n}_2(x,y) (E_x^2 + E_y^2 + E_z^2) {\mathbf E}.
\ea
Here, $\bar{n}_2(x,y)$ stands for the nonlinear index, which as it indicates, may depend on location.
Other models of light-matter interactions that have been used in simulations are described in articles on 
filamentation \cite{CouaironPhysRep07,BergeRPP07} and Ref. \cite{Guide11} shows methods for their numerical 
implementation.

To keep notation simple, we will not use current density explicitly. 
In general, nonlinear interactions with the medium can be equivalently formulated either in the polarization or 
current density language, so this means no loss of generality. 
In numerical simulations, using both current and polarization may actually be convenient, 
and it only requires trivial extension of our results.

\subsection{Fields in terms of analytic signals}

In numerical simulations, it is often easier to work with analytic signals of the electric field.
Here we use analytic signals to represent all real quantities (${\mathbf E}(t),{\mathbf P}(t)$).
For example, the electric field is obtained as a real part of its analytic signal:
\be
{\mathbf E} = \mbox{Re}\{ \vec E(x,y,z,t) \},
\ee
which has its spectrum restricted to positive frequencies:
\be
\vec E(x,y,z,t) = \int_0^\infty \!\!\! d\omega \vec E(x,y,z,\omega) e^{-i \omega t}.
\ee
Here, and in what follows, we will distinguish between temporal and spectral representations of
functions through their respective arguments $t$ and $\omega$.
Because the only time we need the representation of the electric field in the time domain is when we compute 
the nonlinear medium response (i.e., polarization), we will work mostly in the spectral representation.

\subsection{Derivation of Directional Pulse Propagation Equations}

Our departure point is the wave equation for the electric field,
accompanied by a constraint in the form of the divergence equation. 
\be
\nabla \nabla \cdot \vec E  - \nabla^2 \vec E =  
\frac{\omega^2}{c^2} \left(\epsilon \vec E  + \frac{1}{\epsilon_0} \vec P\right)
\  , \  \nabla \cdot \vec D = 0
\ee
The divergence equation deserves a note. 
While we have assumed no free charges and currents,
high intensities can lead to medium ionization and subsequently to electrons 
drifting away from their parent ions. 
However, our treatment aims to describe femtosecond pulses. 
On such a short time scale we can safely assume that even when ionization occurs, 
the positive and negative charges do not have enough time to separate, 
and the average local charge remains zero.
Therefore, using the divergence equation, 
\be
\nabla \cdot \vec D = \epsilon_0 \epsilon\nabla\cdot \vec E 
+ \epsilon_0 \vec E \cdot\nabla\epsilon + \nabla\cdot\vec P  = 0,
\label{eq:div}
\ee
$\nabla \cdot \vec E$ can be expressed in terms of the nonlinear polarization divergence and
transverse electric field components as follows
\be
-\nabla\cdot\vec E=\frac{1}{\epsilon}\vec E_\perp \cdot\nabla_\perp \epsilon
+\frac{1}{\epsilon_0\epsilon}\nabla\cdot\vec P
\ee
The transverse ($x,y$) part of the wave equation is thus rewritten to separate the linear and nonlinear terms
\be
-\partial_{zz}\vec E_\perp = \hat L\vec E_\perp + \hat N_\perp[ \vec E ]. \label{eq:waveperp}
\ee
Here, the linear operator $\hat L$ is related to the corresponding Helmholtz equation 
(for a fixed angular frequency). 
It acts only on the transverse electric field vector 
\be
 \hat L \vec E_\perp \equiv   
  \frac{\omega^2}{c^2} \epsilon(r_\perp, \omega)\vec E_\perp  
+ \Delta_\perp \vec E_\perp
+ \nabla \frac{1}{\epsilon}\vec E_\perp . \nabla_\perp \epsilon.\label{eq:linearprop}
\ee
The nonlinear operator $\hat N$ acts on $\vec E_\perp$, but in general, also depends on the $E_z$ component
\be
\hat N[ \vec E ] \equiv 
\frac{\omega^2}{\epsilon_0 c^2} \vec P(\vec E)
+ \nabla \frac{1}{\epsilon_0\epsilon} \nabla \cdot \vec P(\vec E).
\ee
We will address how to obtain $E_z$ later. 
For now, let us assume that it can be calculated once $E_{x,y}$ are known. 
To obtain propagation equations for $\vec E_\perp(z,x,y,\omega)$, we first introduce auxiliary field amplitudes, 
effectively doubling the number of variables used to describe the electric field:
\be
E_i(z,x,y,\omega) = E_i^+(z,x,y,\omega) + E_i^-(z,x,y,\omega), \label{eq:Epm}
\ee
with
\ba
E_i^+ &=& A^+_i (z,x,y,\omega) e^{ + i \zeta z} \nonumber\\
E_i^- &=& A^-_i (z,x,y,\omega) e^{ - i \zeta z},
\ea
where $i = x,y$ and $\zeta$ stands for a parameter to be chosen freely. 
Clearly, since no $\zeta$ appears in Maxwell's nor the wave equations, physical observables must not depend on 
the concrete choice of $\zeta$, and this ``gauge invariance'' will become manifest when we arrive at our final 
result.
We will term $\zeta$ a reference wavenumber to emphasize the fact that it has no physical meaning by itself.

It is also important to keep in mind that the positive and negative wavenumber parts $E_i^\pm$ of the 
field are, in general, not the forward and backward propagating portions of the total waveform.
So far we have not restricted how fast $A^\pm_i$ can change with $z$.
In principle, they could evolve so fast that their variation would completely override the exponential 
factors $e^{ \pm i \zeta z}$ accompanying them.  
That is why both $E_i^\pm$ can contribute to waves propagating in the positive and negative $z$ direction
(see Ref. \cite{kinsler_unidirectional_2010} for how this occurs).

By representing a single function $E_i(z,x,y,\omega)$ as a combination of two functions $E_i^\pm(z,x,y,\omega)$,
we have added artificial degrees of freedom. 
These will be taken back by requiring that  $E_i^\pm$ satisfy a relation of our choice. 
Concretely, we impose an additional constraint in the form
\be
 e^{ + i \zeta z} \partial_z A^+_i (z,x,y,\omega) + e^{ - i \zeta z} \partial_z A^-_i (z,x,y,\omega)  = 0
\label{eq:constraint}
\ee
The rationale behind this constraint is exactly the same as in the variation of constants method for 
differential equations. 
Namely, this representation eliminates the second derivatives when one evaluates $\partial_{zz} E$. 
Because of the constraint, the first derivative simplifies to
\be
\partial_z E_i = i \zeta (E_i^+ - E_i^-)
\label{eq:derz}
\ee
and the second derivative to
\be
\partial_{zz} E_i = -\zeta^2 E_i 
     +i \zeta e^{+ i \zeta z} \partial_z A^+_i
     -i \zeta e^{- i \zeta z} \partial_z A^-_i
\ee
Using this in the wave equation (\ref{eq:waveperp}) together with the constraint of Eq. (\ref{eq:constraint}), 
we obtain the evolution equations for the auxiliary amplitudes $A^\pm$
\be
\partial_z A^\pm_i  =
\frac{\pm i}{2 \zeta}  e^{ \mp i \zeta z} \left[ (\hat L - \zeta^2)\vec E_\perp + \hat N[\vec E] \right].
\ee
To evaluate the right hand side of this system, $\vec E_\perp$ is expressed in terms of $A^\pm_i$,
and $E_z$ is subsequently obtained from the $z$-component of the wave equation.
Using Eq. (\ref{eq:derz}), the latter can be written as follows
\ba
&&i\zeta\partial_x(E_x^+ - E_x^-) + i\zeta\partial_y(E_y^+ - E_y^-)-\frac{\omega^2}{c^2 \epsilon_0} P_z(\vec E)\nonumber\\
&& = \partial_{xx} E_z + \partial_{yy} E_z + \frac{ \omega^2 \epsilon}{c^2} E_z. \label{eq:EZ}
\ea
If not for nonlinearity, this is an inhomogeneous Helmholtz equation that determines $E_z$
in terms of the transverse field components.
The part of the polarization component $P_z$ which is nonlinear in $E_z$, is usually very small
and therefore normally neglected. 
Should one not be satisfied with such an approximation, the above equation can be solved by iteration. 
For example, in Kerr media for intensities typical of femtosecond filaments, 
a single iteration already gives an accurate result.

The next step consists in identifying the parts of the electric field waveform which
propagate in the positive and negative directions along the $z$ axis. 
The resulting equations become more intuitive when expressed in terms of auxiliary $E$-fields:
\ba
\partial_z E_i^+ &=& + i \zeta E_i^+ + \frac{i}{2 \zeta} \left[ (\hat L - \zeta^2)\vec E_\perp + \hat N[\vec E] \right]\\
\partial_z E_i^- &=& - i \zeta E_i^- - \frac{i}{2 \zeta} \left[ (\hat L - \zeta^2)\vec E_\perp + \hat N[\vec E] \right]
\ea
This system, completed by Eq. (\ref{eq:EZ}), is equivalent to the wave equation with the divergence constraint, 
and can be solved in principle. 
However, in this form, it poses two problems.
First, in general, it would require very short propagation steps in order to resolve both 
the forward and backward propagating waves. 
Second, the physical input conditions for simulations are normally given such that the problem to
solve is a boundary value problem rather than an initial value problem. 
The latter point becomes evident when we realize that it is only the forward-propagating field component 
that is specified at $z=0$ (e.g., at the laser output). 
The second condition is that the backward propagating field is zero at $z\to\infty$
(i.e., at the far end of a laboratory). 
Such a boundary value problem would be rather difficult to solve.
Fortunately, in many cases the backward propagating wave can be neglected. 
We shall therefore derive the beam propagation equations that account for such a situation, but
in the process shall identify the forward and backward propagating fields 
(and see that they are, in general, different from $E^\pm$).

In matrix notation, the propagation equations read
\ba
\partial_z
\begin{pmatrix}
 \vec E_{\perp}^+ \\
 \vec E_{\perp}^- \\
\end{pmatrix}
&=& i 
\begin{pmatrix}
   \zeta + \frac{\hat L  - \zeta^2}{2 \zeta} & \phantom{-\zeta}+ \frac{\hat L  - \zeta^2}{2 \zeta} \\
   \phantom{\zeta}  - \frac{\hat L  - \zeta^2}{2 \zeta}       & -\zeta - \frac{\hat L  - \zeta^2}{2 \zeta}
\end{pmatrix}
\begin{pmatrix}
   \vec E_\perp^+ \\
   \vec E_\perp^-
\end{pmatrix} \nonumber\\
&& + \frac{i}{2\zeta} 
\begin{pmatrix}
  +\hat N_\perp[\vec E]\\
  -\hat N_\perp[\vec E]
\end{pmatrix}.
\label{eq:M}
\ea
Having separated the linear and nonlinear part of the evolution operator, we are in the position
to determine the forward and backward propagating parts of the field. 
This division will be defined with respect to the linear system. 
In the spirit similar to Ref.~\cite{KolesikPRL02}, 
two projector operators can be constructed from the Helmholtz operator 
$\hat L$ and its square root $\hat L^{\frac{1}{2}}$:
\ba
{\cal P}_F
\equiv \frac{\hat L^{-\frac{1}{2}}}{4 \zeta}
\begin{pmatrix}
   +(\zeta + \hat L^{\frac{1}{2}})^2 & +(\hat L - \zeta^2) \\
   - (\hat L - \zeta^2) & -(\zeta - \hat L^{\frac{1}{2}})^2
\end{pmatrix}
\\
{\cal P}_B
\equiv \frac{\hat L^{-\frac{1}{2}}}{4 \zeta}
\begin{pmatrix}
  -(\zeta - \hat L^{\frac{1}{2}})^2 & -(\hat L - \zeta^2) \\
  +(\hat L - \zeta^2) &  +(\zeta + \hat L^{\frac{1}{2}})^2 
\end{pmatrix}
\label{eq:projs}
\ea
It is straightforward to show that these operators have the expected properties of projectors, 
in particular they are idempotent
\be
{\cal P}_F^2 = {\cal P}_F \ \ \ \ {\cal P}_B^2 = {\cal P}_B \label{eq:idempotent}
\ee
and they constitute a unity decomposition
\be
{\cal P}_F  +  {\cal P}_B = 1  \  , \  {\cal P}_F  {\cal P}_B = {\cal P}_F  {\cal P}_B = 0. \label{eq:unity}
\ee
These projectors also commute with the linear evolution operator in Eq. (\ref{eq:M}), and direct calculation 
shows that their eigenvectors have propagation constants corresponding to forward and backward modes propagating
in the linear system. 
Thus, we can use these projectors to obtain the true forward and backward propagating field components. 
If the total field is given in terms of the auxiliary amplitudes $\vec E_\perp^\pm$, 
then the forward portion of the wave is obtained as
\ba
E_F &=&
\begin{pmatrix}
  1 & 1
\end{pmatrix}
{\cal P}_F
\begin{pmatrix}
  \vec E_\perp^+ \\
  \vec E_\perp^-
\end{pmatrix}\nonumber\\
&=& \frac{1}{2}\left[(\vec E_\perp^+ + \vec E_\perp^-) + \hat L^{-\frac{1}{2}} \zeta ( \vec E_\perp^+ - \vec E_\perp^-)\right].
\ea
This expression contains the reference wavenumber, which might suggest that it depends on our artificial 
split of the field in Eq. (\ref{eq:Epm}). 
However, because of Eq. (\ref{eq:derz}), the above expression reduces to
\be
E_F = \frac{1}{2}\vec E_\perp - \frac{i}{2} \hat L^{-\frac{1}{2}} \partial_z \vec E_\perp
\ee
and the backward propagating component is obtained as
\be
E_B = \frac{1}{2}\vec E_\perp + \frac{i}{2} \hat L^{-\frac{1}{2}} \partial_z \vec E_\perp .
\ee
As they must be, these forward/backward amplitudes are independent of the reference wavenumber.
Our aim is to express the pulse evolution equations in term of these directional fields. 
Because the projector operators are $z$-independent, the simplest way is to apply them directly to
the propagation equations. 
In other words, we need to compute
\ba
\partial_z E^F_\perp &=& 
\begin{pmatrix}
  1 & 1
\end{pmatrix}
{\cal P}_F
\begin{pmatrix}
  \partial_z \vec E_\perp^+ \\
  \partial_z \vec E_\perp^-
\end{pmatrix} \\
\partial_z E^B_\perp &=&
\begin{pmatrix}
  1 & 1
\end{pmatrix}
{\cal P}_B
\begin{pmatrix}
  \partial_z \vec E_\perp^+\\
  \partial_z \vec E_\perp^-
\end{pmatrix}.
\ea
Inserting the right hand side from Eq. (\ref{eq:M}) and using the projector properties of Eqs. 
(\ref{eq:idempotent}) and (\ref{eq:unity}), we obtain a pair of coupled equations for the forward 
and backward fields 
\ba
\partial_z E^F_\perp &=& +i \sqrt{\hat L} E^F_\perp + \frac{i}{2\sqrt{\hat L}} \hat N_\perp[E^F + E^B] \cr
\partial_z E^B_\perp &=& -i \sqrt{\hat L} E^B_\perp - \frac{i}{2\sqrt{\hat L}} \hat N_\perp[E^F + E^B]
\label{eq:EFB}
\ea
This is a generalization of the coupled pair of Unidirectional Pulse Propagation Equations 
(\ref{eq:z1}) and (\ref{eq:z2}).
We show later that in a homogeneous medium, for which we have an explicit expression for the 
square root of the Helmholtz operator, these equations reduce to UPPEs as they should.

Similar to the case for bulk media, Eq. (\ref{eq:EFB}) is not the best for numerical implementation.
We transform this system into a form analogous to that of bulk UPPE (\ref{eq:As}) such that we can adopt
the same numerical solution strategy. 
Toward this purpose, we use amplitudes which will exhibit evolution only if some nonlinearity is present:
\be
E^F_\perp = e^{+i \sqrt{\hat L} z} A^F_\perp (z) \ , \  E^B_\perp = e^{-i \sqrt{\hat L} z} A^B_\perp (z) 
\label{eq:EvsA}
\ee
In this representation, Eq. (\ref{eq:EFB}) reads
\ba
\partial_z A^F_\perp &=& \frac{+i}{2\sqrt{\hat L}}e^{-i \sqrt{\hat L} z}\hat N_\perp[ e^{+i \sqrt{\hat L} z}A^F + e^{-i \sqrt{\hat L} z} A^B] \nonumber\\
\partial_z A^B_\perp &=& \frac{-i}{2\sqrt{\hat L}}e^{+i \sqrt{\hat L} z}\hat N_\perp[ e^{+i \sqrt{\hat L} z}A^F + e^{-i \sqrt{\hat L} z} A^B]\nonumber\\
&&\phantom{XXX}
\ea
This shows explicitly that the forward and backward propagating waves are mutually coupled in the nonlinear terms.
It is obvious that for strong nonlinearity, our forward-backward projection loses its intended meaning, 
because it can renormalize and couple waves propagating in both main 
directions \cite{kinsler_limits_2007}. 
Thus, we arrive at a point where we must adopt an approximation which will allow us to reduce the full system 
to a single unidirectional equation.

\subsection{Unidirectional propagation approximation}

Our final step is to adopt the unidirectional approximation, where we assume that the nature and strength of
nonlinearity is such that only negligible backward propagating fields are generated. 
Then, the nonlinear term can be approximated as
\be
 \hat N_\perp[ e^{+i \sqrt{\hat L} z}  A^F + e^{-i \sqrt{\hat L} z} A^B] \approx  \hat N_\perp[ e^{+i \sqrt{\hat L} z}  A^F]
\ee
and the system can be restricted to only the forward-propagating field:
\be
\partial_z A^F_\perp(r_\perp,\omega,z) = + \frac{i}{2\sqrt{\hat L}} e^{-i \sqrt{\hat L} z} \hat N_\perp[ e^{+i \sqrt{\hat L} z}  A^F]. \label{eq:general}
\ee
This is the sought-after generalization of the Unidirectional Pulse Propagation Equation. 
As expected, the structure of this system is completely analogous to the bulk UPPEs of Eq. (\ref{eq:As}), with 
the exception that the linear propagator is formally expressed in terms of a Helmholtz square root operator, 
instead of plane-wave expansion.
The most pronounced difference is that Eq. (\ref{eq:general}) is natively represented in the 
mixed representation.
It retains the spectral treatment of the time dimension and with that, it preserves the
ability to treat chromatic and nonlinear properties of the material exactly. 
On the other hand, the transverse dimensions are represented in real space, which is the natural choice for the 
implementation of the linear propagator in a structured medium with strong refractive index variations.

\subsection{Special case: reduction to UPPE in a homogeneous medium}

Before going into how this pulse evolution equation can be solved numerically, let us illustrate how
it reduces to the well-known bulk UPPE for a homogeneous medium. 
First, we recall that for a homogeneous medium, we know that plane waves are eigenfunctions of the Helmholtz 
operator $\hat L$ and that in the plane-wave representation, the linear propagator reduces to
multiplication by a phase factor given by the propagation constant $k_z(\omega, k_\perp)$:  
\[
e^{-i \sqrt{\hat L} z} = e^{-i k_z(\omega, k_\perp) z} .
\]
It is therefore sufficient to Fourier-transform Eq. (\ref{eq:general}) from the ($x,y$) space to 
the transverse wavenumber space ($k_x,k_y$) to obtain
\be
\partial_z A^F_\perp(k_\perp,\omega,z) = + \frac{i}{2 k_z} e^{-i k_z z} \hat N_\perp[ e^{+i k_z z}  A^F],
\ee
then use Eq. (\ref{eq:EvsA})
\be
\partial_z  E^F_\perp (k_\perp,\omega,z) = +i k_z E^F_\perp (k_\perp,\omega,z) 
+  \frac{i}{2 k_z} \hat N_\perp[ E^F],
\ee
and express $\hat N$ in terms of polarization:
\be
\partial_z E^F = i k_z E^F
+  \frac{i}{2 k_z} \left[ \frac{\omega^2}{\epsilon_0 c^2} \vec P(\vec E)
-  \frac{1}{\epsilon_0\epsilon} \vec k \vec k \cdot \vec P(\vec E)
\right].
\label{eq:aux1}
\ee
Only the transverse components in this equation constitute the evolution system, but in this full-vector form, 
it is easy to see that the operator acting on the polarization term produces the transverse part of the 
nonlinear response, namely
\be
\left[ \frac{\omega^2}{\epsilon_0 c^2} \vec P(\vec E)
-  \frac{1}{\epsilon_0\epsilon} \vec k \vec k \cdot \vec P(\vec E) \right]
= \frac{ \omega^2}{\epsilon_0 c^2 } \left[ 1 - \frac{\vec k \vec k \cdot}{k^2 } \right] \vec P(\vec E).
\label{eq:proj}
\ee
The projector operator in the square brackets can be replaced by a sum over projectors on
the polarization vectors $\vec e_s$
\be
\left[ 1 - \frac{\vec k \vec k \cdot}{k^2 } \right] = \sum_s \vec e_s \vec e_s \cdot
\ee
Using this in Eq. (\ref{eq:proj}) and inserting it into Eq. (\ref{eq:aux1}), we obtain
\be
\partial_z  E^F_\perp(z,\omega,k_\perp) = i k_z  E^F_\perp
+  \frac{i\omega^2}{2 \epsilon_0 c^2 k_z} \sum_s \vec e_s^\perp \vec e_s \cdot \vec P(\vec E),
\ee
which is identical to the homogeneous medium UPPE of Eq. (\ref{eq:z1}) (with the current density term omitted).
Thus, as it must, the generalized pulse propagation equation (\ref{eq:general}) passes this sanity 
check and reduces to the UPPE if the medium is homogeneous.

\section{Numerical solution strategy\label{sec:numeric}}

In this section we sketch, in broad strokes, an approach for the numerical solution. 
It builds on the ODE-based method for solving UPPE systems and combines it with a
wide-angle beam-propagation solver used to evaluate the linear propagator $\exp(i \hat L^{1/2} z)$.

The core of the pulse propagator of Eq. (\ref{eq:general}) is an ODE system, with $z$ being the independent
variable.
The equation is evaluated at every transverse spatial location $r_\perp$ and frequency $\omega$ while 
being incremented along the propagation direction $z$. 
During a single ODE step, the right hand side of Eq. (\ref{eq:general}) has to be evaluated multiple times
at different values of $z$ which are subject to the choice of the specific ODE algorithm.
Because the integration is normally executed with an adaptive integration step, one cannot determine
beforehand at what specific $z$ locations the term [$\exp(i \hat L^{1/2} z) A$] needs to be computed --
an algorithm is needed to evaluate the right hand side for \emph{any} small value of $z$. 
For the ODE solver, we use the open source Gnu Scientific Library (GSL), 
but any implementation with the following capabilities can be chosen. 
One necessary feature of a suitable ODE library is a driver for adaptive step control, 
with a robust algorithm monitoring the accuracy of the numerical solution. 
Another necessity is that the library contains methods which do not require Jacobian evaluation, 
because such methods are not suitable for UPPE-like ODE systems \cite{Guide11}.
We typically employ the Runge-Kutta-Fehlberg method, however, another useful ODE library feature is the 
capability to switch easily between different integration methods.
Regarding its structure and method of solution as an ODE system, the generalized propagation equation does not differ from an 
ordinary UPPE.  What is different is the implementation of the linear propagator. 

Because the linear propagator is diagonal in angular frequency, this task is equivalent to a set
of uncoupled beam-propagation problems. 
In other words, the action of $\exp(i \hat L^{1/2} z)\psi$ only requires one independent BPM-like update 
for each $\omega$ resolved in the simulation. 
This portion of the algorithm is therefore ``embarrassingly parallel,'' with perfect
balance and no inter-dependencies between calculations performed for different angular frequencies. 
There are many wide-angle BPM methods available, and any of them can be utilized, in principle.
For instance, one can evaluate the linear propagator by a Pad\'e approximant. 
Defining $\beta^2(\omega) \equiv \omega^2\epsilon(\omega) / c^2$, the dominant part of the Helmholtz operator, 
one writes
\be
e^{i \sqrt{\hat L} \Delta z}=e^{i \beta \sqrt{1 + \hat X} \Delta z} = \prod_k \frac{\hat X + a_k}{\hat X + b_k}.
\ee
The coefficients $a_k,b_k$ depend on $\Delta z$ and are chosen as to reproduce the Taylor 
expansion of the left hand side. 
For example,
\be
\frac{4 i + (i - \beta \Delta z)\hat X }{4 i + (i + \beta \Delta z)\hat X } e^{i\beta \Delta z}
\label{eq:padeapp}
\ee
is second-order accurate in $\hat X$ with the error scaling as $\sim\beta \Delta z \hat X^3$.
Various higher order approximations can be constructed in the same spirit. If the operator
$X$ acts in a non-trivial way along both spatial dimensions $x,y$, it is often further split into
``one-dimensional'' components so that the resulting matrices are band-diagonal.

Similar techniques can be used to compute the inverse square root of $L$ that acts on the
nonlinear response term in Eq. (\ref{eq:general}). 
However, this operator can be approximated by $L^{1/2}\approx \omega n(\omega) / c$ as is usually done in
filamentation simulations \cite{Guide11}. 
This is sufficient unless the spatial profile of the nonlinear polarization 
becomes ``focused'' to wavelength scale. 

With the linear propagator implemented as a ``BPM-based plug-in,'' 
the solution proceeds in steps with two stages:
\begin{enumerate}
\item \emph{Call the ODE solver}.
One integration step is executed that updates the current $A(r_\perp,\omega,z)$
into the new $A(r_\perp,\omega,z + \Delta z)$. 
The ODE solver algorithm invokes computation of the right hand side of 
Eq. (\ref{eq:general})
\[
 + \frac{i}{2 \beta} e^{-i \sqrt{\hat L} \delta z} \hat N_\perp \left[e^{+i \sqrt{\hat L} \delta z}A^F\right],
\]
which contains two applications of the linear propagator [e.g., Eq. (\ref{eq:padeapp})] for a sub-step 
$\pm\delta z$.
Behind the scenes, the solver determines the maximum step $\Delta z$ possible on a global scale, 
since some parts of the grid may contain finer features, and require shorter integration steps than others. 
Unlike the fully spectral UPPE, the length of the integration step the ODE solver is permitted to take is 
bounded from above by the maximum step allowed by the BPM method used for the linear propagator. 

\item \emph{Re-align the spectral amplitudes.}
The point along $z$ at which $A$ and $E$ amplitudes coincide can, of course, be chosen arbitrarily. 
It is advantageous to renew this synchronization point after each ODE step such that $A$ and $E$ 
coincide at the beginning of the ODE step. 
This is achieved by 
\be
A^F_\perp \leftarrow  e^{+i \sqrt{\hat L} \Delta z} A^F_\perp ,
\label{eq:renew}
\ee
which amounts to yet another application of the linear-problem propagator to the current solution. 
Naturally, $\Delta z$ must be obtained from the ODE solver as the actual length of the last adaptive 
integration step. 
This repeated re-alignment step is normally implemented in bulk-media UPPE solvers as well, 
but there implementations without it are possible, in principle.
Here, it is crucial that the step length in the linear propagator is kept small, and application of 
Eq. (\ref{eq:renew}) ensures that $\delta z$ is always smaller than the maximal step allowed in the ODE solver.

\end{enumerate}

In a nutshell, the above procedure describes the standard UPPE solution method modified
in two ways: First, a BPM-based propagator is utilized for the (short-step) linear advancement
of the optical field, and second, real-space representation of transverse dimensions
is retained at all times.

\section{Summary\label{sec:sum}}

We have presented a generalization of the Unidirectional Pulse Propagation Equation suitable for structures 
characterized by material interfaces parallel to the pulse propagation direction and by strong differences 
between the properties of the constituent materials. 
While the main result of Eq. (\ref{eq:general}) is somewhat intuitive, 
we show a rigorous derivation based on identification of the forward and backward propagating wave components. 
These are expressed in terms of projection operators [Eq. (\ref{eq:projs})] akin to those we have previously 
used in bulk media~\cite{Kolesik2002}. 
They allow expression of the generalized UPPE in terms of the linear propagator,
and they ``isolate'' the nonlinear interactions with the medium, such that the evolution is described in terms 
of spectral amplitudes which only evolve due to non-zero nonlinearity. 

The generalized UPPE uses a mixed representation: spectral for the time/frequency dimension and 
real-space for the transverse (to the direction of propagation) dimensions. 
The linear propagator can be based on one of the many available beam-propagation methods. 
The concrete choice of the method will depend on the given geometry. 
For example, the important case of nearly plane-parallel wave guide structures \cite{Akturk:09,Arnold:09} can be 
treated by a one-dimensional WA-BPM combined with the plane-wave expansion in the free-propagating direction. 
Independently of the chosen BPM approach, the numerical solution strategy developed previously
for bulk-media UPPEs can be used with relatively minor modifications.

There is an increasing interest in extreme nonlinear optics confined to wave-guiding structures of different kinds.
It is therefore expected that our results will find application in various implementations of efficient pulse 
propagation solvers, especially
situations in which both the geometry of the structure and waveform reshaping due to nonlinear interactions 
play important roles.

\begin{acknowledgments}
  This work was supported by the U.S. Air Force Office for Scientific
  Research under grant FA9550-11-1-0144.
\end{acknowledgments}


\end{document}